\documentclass{epl}
\usepackage{amsmath}
\usepackage{amssymb}
\usepackage{graphicx}
\usepackage{color}

\newcommand{\dd}{\mathrm{d}}
\newcommand{\mean}[1]{\left\langle #1 \right\rangle}

\newcommand{\pd}[2]{\frac{\partial #1}{\partial #2}}
\newcommand{\fd}[2]{\frac{\delta #1}{\delta #2}}

\newcommand{\IInt}[3]{\int_{#2}^{#3}\dd #1\;}
\newcommand{\Path}[1]{\int[\dd #1]\;}

\newcommand{\ps}{p_\mathrm{s}}
\newcommand{\js}{j_\mathrm{s}}
\newcommand{\vs}{\nu_\mathrm{s}}
\newcommand{\vm}{v_\mathrm{s}}
\newcommand{\Deff}{D_\mathrm{eff}}
\newcommand{\Teff}{T_\mathrm{eff}}
\newcommand{\Req}{R^\mathrm{eq}}

\newcommand{\fpr}{{f_\mathrm{p}}}
\newcommand{\p}{_\mathrm{p}}
\newcommand{\e}{_\mathrm{eq}}
\newcommand{\tot}{_\mathrm{tot}}
\newcommand{\hk}{_\mathrm{hk}}
\newcommand{\A}{\mathcal A}
\newcommand{\PP}{\mathcal P}

\newcommand{\ness}{nonequilibrium steady state}
\newcommand{\fdt}{FDT}
\newcommand{\fdr}{fluctuation-dissipation relation}

\begin{document}

\title{Restoring a fluctuation-dissipation theorem in a nonequilibrium steady
  state}
\shorttitle{Fluctuation-dissipation theorem in nonequilibrium}
\author{T.~Speck \and U.~Seifert}
\shortauthor{T.~Speck \etal}
\institute{{II.} Institut f\"ur Theoretische Physik, Universit\"at Stuttgart,
70550 Stuttgart, Germany}

\maketitle

\begin{abstract}
  In a \ness, the violation of the fluctuation-dissipation theorem (FDT) is
  connected to breaking detailed balance. For the velocity correlations of a
  driven colloidal particle we calculate an explicit expression of the FDT
  violation. The equilibrium form of the FDT can be restored by measuring the
  velocity with respect to the {\em local} mean velocity.
\end{abstract}


\section{Introduction}

The fluctuation-dissipation theorem (FDT) relates the correlation function of
thermally driven fluctuations in equilibrium with the response of the system
to a small external perturbation~\cite{kubo}. Loosely speaking following
Onsager, the decay of a fluctuation is independent of whether it has been
created spontaneously due to thermal noise or whether is has been induced by a
small applied force. According to common wisdom, the \fdt\ breaks down in a
nonequilibrium system. Indeed, a large body of studies deals with violations
of the \fdt\ in nonequilibrium, in particular, for aging systems (for reviews,
see refs.~\cite{cris03,cala05}), fluids and colloidal suspensions driven by
shear flow~\cite{barr01,fuch05}, and biophysical systems (for an example, see
ref.~\cite{mart01}).

An important aspect of \ness s is that detailed balance is broken, which
requires permanent dissipation of energy. In the case of driven colloidal
systems, the heat thus dissipated has been identified as ``housekeeping
heat''~\cite{oono98,hata01,spec05a}. Since both the housekeeping heat and the
violation of the FDT originate from breaking detailed balance, a deeper
connection between these two concepts can be expected~\cite{hara05}. Since
furthermore the velocity is the crucial quantity entering the housekeeping
heat, this connection should become apparent in the violation of the velocity
FDT.

In this Letter, we first derive an explicit expression for this violation. We
then demonstrate the restoration of the usual equilibrium form of the velocity
\fdt\ if the velocity fluctuations are properly measured with respect to the
{\em local} mean velocity rather than with respect to the {\em global} mean
velocity. Our result thus suggests that the decay of fluctuations around the
steady state is the same, whether they are spontaneously created or externally
induced.


\section{The system}

\begin{figure}[t]
  \onefigure[width=5cm]{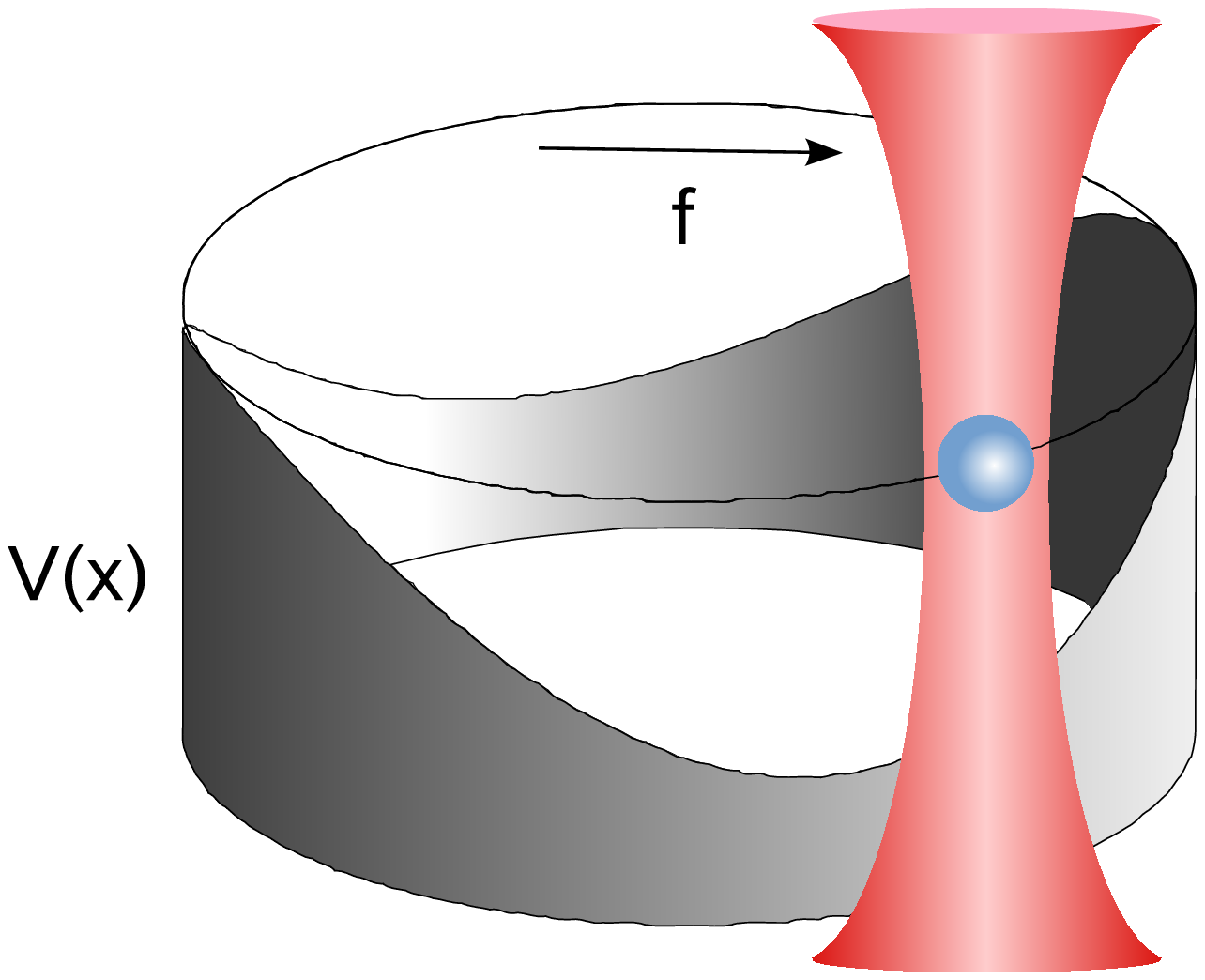}
  \caption{A colloidal particle moving in a periodic potential $V(x)$ with
    periodicity $l=2\pi$ and angular position $x$. A force $f$ can be applied
    directly to the particle, driving it into a \ness.}
  \label{fig:ring}
\end{figure}

As a paradigm, we study one-dimensional systems with periodicity $l$ subject
to stochastic dynamics, where the periodicity is necessary to eventually reach
a \ness~(see fig.~\ref{fig:ring}). Overdamped diffusive motion is then
governed by the Langevin equation~\cite{risken}
\begin{equation}
  \label{eq:lang}
  \dot x(t) = \mu\left[F(x)+\fpr(t)\right] + \eta(t).
\end{equation}
The total force $F(x)=-V'(x)+f$ can be split into the gradient of a periodic
potential $V(x+l)=V(x)$ and a non-conservative force $f$, where the prime
denotes the derivative with respect to $x$. The force $\fpr$ represents a
small external perturbation. The thermal noise $\eta(t)$ has zero mean and
correlations $\mean{\eta(t)\eta(\tau)}=2D\delta(t-\tau)$. In equilibrium, bare
mobility $\mu$, temperature $T$, and diffusion coefficient $D$ are connected
by the Einstein relation $D=\mu T$ with Boltzmann's constant set to $1$
throughout the paper. We will keep the Einstein relation despite the fact that
the system is driven into a \ness\ for $f\neq0$, which expresses the
assumption that the driving does not affect the heat bath. The brackets
$\mean{\cdot}$ represent the average over the thermal noise.

An unperturbed ($\fpr=0$) one-dimensional system in a \ness\ is distinguished
from equilibrium by a nonzero, constant current
\begin{equation}
  \label{eq:js}
  \js = \mu F(x)\ps(x) - D\ps'(x) \neq 0.
\end{equation}
We define the stationary distribution $\ps(x)\equiv\exp[-\phi(x)]$ introducing
the generalized potential $\phi(x)$. In addition, the notion of a local mean
velocity
\begin{equation}
  \label{eq:vs}
  \vs(x) \equiv \js/\ps(x) = \mu F(x) + D\phi'(x)
\end{equation}
will become crucial. It is given by the drift velocity $\mu F(x)$ plus the
gradient of the potential $\phi(x)$. The local mean velocity $\vs(x)$ can be
regarded as a measure of the local violation of detailed balance. After
averaging
\begin{equation}
  \vm \equiv \mean{\vs} = \IInt{x}{0}{l}\vs(x)\ps(x)
  = \mu \mean{F} - D[\ps]^l_0 = \mu \mean{F} = \mean{\dot x}
\end{equation}
it becomes the global mean velocity of the particle in a steady state as
expected. The boundary term vanishes due to the periodicity of the system.

\section{Fluctuation-dissipation relations}

A small probing force $\fpr(t)$ is applied after the system has been prepared
in a steady state described by the stationary probability $\ps(x)$. The linear
response of any observable $A$ to this perturbation is defined as
\begin{equation}
  \label{eq:response}
  R_A(t-\tau) \equiv 
  \left.\frac{\delta\mean{A(x(t))}\p}{\delta\fpr(\tau)}\right|_{\fpr=0}
  \quad(t\geqslant\tau),
\end{equation}
which depends only on the time difference $t-\tau$. Due to causality the
response function is zero for $t<\tau$ and in what follows we always take
$t\geqslant\tau$. We distinguish averages $\mean{\cdot}\e$ in the equilibrium
system ($f=0$), averages $\mean{\cdot}$ in the unperturbed system in a \ness,
and averages $\mean{\cdot}\p$ in the perturbed system ($\fpr\neq0$).

The various \fdr s state that the response function of an observable
is not independent of its correlations. In particular, in equilibrium the
well-known \fdt
\begin{equation}
  \label{eq:fdt:eq}
  T\Req_A(t-\tau) = \partial_\tau\mean{A(x(t))B(x(\tau))}\e
\end{equation}
relates the response function of $A$ to the correlations of this observable
with a unique observable $B$ conjugated to $\fpr$ with respect to energy.
However, even in a \ness\ the less known \fdr
\begin{equation}
  \label{eq:fdt:noneq}
  TR_A(t-\tau) = \mean{A(x(t))D\phi'(x(\tau))}
\end{equation}
correlates the observable $A$ to the gradient of the generalized potential
$\phi(x)$~\cite{risken}. Similar relations have been discussed also for
chaotic systems~\cite{falc95}. A third relation involving the noise can be
derived from the Gaussian weight
\begin{equation}
  \PP[\eta] \sim \exp\left\{ -\frac{1}{4D}\IInt{\tau}{t_0}{t}\eta^2(\tau)
  \right\}
\end{equation}
of the noise trajectory $\eta(\tau)$ in the time interval
$t_0\leqslant\tau\leqslant t$. Since $A(x(t))$ can formally be expressed as a
functional $\A[\eta]$ of the noise history and variation with respect to
$\mu\fpr(\tau)$ is equivalent to a variation of $\eta(\tau)$ in
eq.~\eqref{eq:lang}, we have
\begin{equation}
  \label{eq:fdt:noise}
  TR_A(t-\tau) = D\mean{\fd{\A[\eta]}{\eta(\tau)}}
  = D\Path{\eta}\fd{\A[\eta]}{\eta(\tau)}\PP[\eta]
  = -D\Path{\eta}\A[\eta]\fd{\PP[\eta]}{\eta(\tau)}
  = \frac{1}{2}\mean{A(x(t))\eta(\tau)}
\end{equation}
through functional integration by parts~\cite{cala05,cugl94}.

\section{Velocity \fdr s}

Because we are interested in the velocity, a \fdr\ involving $\dot x$ instead
of the potential $\phi(x)$ in eq.~\eqref{eq:fdt:noneq} is called for. To this
end we insert the Langevin equation~\eqref{eq:lang} with the force replaced by
the local mean velocity~\eqref{eq:vs} into eq.~\eqref{eq:fdt:noneq}. In the
resulting expression
\begin{equation}
  TR_A(t-\tau) = \mean{A(x(t))\vs(x(\tau))} - \mean{A(x(t))\dot x(\tau)} +
  \mean{A(x(t))\eta(\tau)}
\end{equation}
we use eq.~\eqref{eq:fdt:noise} for the noise term and obtain
\begin{equation}
  \label{eq:fdt:xdot}
  TR_A(t-\tau) = \mean{A(x(t))\dot x(\tau)} - \mean{A(x(t))\vs(x(\tau))}
\end{equation}
holding in a \ness.

First, we consider the equilibrium case where the position $B=x$ and the
probing force $\fpr$ are conjugate. Setting $A=\dot x$ we obtain from
eq.~\eqref{eq:fdt:eq}
\begin{equation}
  \label{eq:1}
  T\Req_{\dot x}(t-\tau) = \mean{\dot x(t)\dot x(\tau)}\e
\end{equation}
the usual form of the FDT for velocities. We can interchange the order of time
derivative and taking the average since the probability distribution implicit
in the brackets is time-independent.

In a \ness, we consequently use eq.~\eqref{eq:fdt:xdot} instead of the
FDT~\eqref{eq:fdt:eq}, leading to
\begin{equation}
  \label{eq:fdt:v}
  TR_{\dot x}(t-\tau) = \mean{\dot x(t)\dot x(\tau)} 
  - \mean{\dot x(t)\vs(x(\tau))}.
\end{equation}
A somewhat related expression has been derived by a different approach in
ref.~\cite{hara05}.  Comparing eq.~\eqref{eq:fdt:v} to the equilibrium
form~\eqref{eq:1} we identify $\mean{\dot x(t)\vs(x(\tau))}$ as the violation
of the \fdt\ and define a normalized violation function
\begin{equation}
  \label{eq:vio}
  I(t-\tau) \equiv \mean{[\dot x(t)-\vm][\vs(x(\tau))-\vm]}
  = \mean{\dot x(t)\vs(x(\tau))} - \vm^2
\end{equation}
such that $I(\infty)=0$.

\section{Restoring the fluctuation-dissipation theorem}

Our crucial observation is that the equilibrium form \eqref{eq:1} can be
restored even in a \ness\ when we consider the {\em relative} velocity
\begin{equation}
  \label{eq:vrel}
  v(t) \equiv \dot x(t) - \vs(x(t)).
\end{equation}
Combining eq.~\eqref{eq:fdt:xdot} and eq.~\eqref{eq:fdt:v}, we obtain
\begin{equation}
  \label{eq:fdt:claim}
  TR_v(t-\tau) = TR_{\dot x}(t-\tau)-TR_{\vs}(t-\tau) = \mean{v(t)v(\tau)}.
\end{equation}
This is the main claim of the present Letter. Physically this relation implies
that fluctuations of the velocity relative to the local mean velocity in a
\ness\ behave like the corresponding response function. In this sense, these
fluctuations cannot be distinguished from equilibrium fluctuations.

\section{Connection to the housekeeping heat}

In a \ness, the heat permanently dissipated in order to maintain the violation
of detailed balance is called the housekeeping heat~\cite{oono98,hata01}. In
the time interval $t_0\leqslant\tau\leqslant t$ it is given as the
functional
\begin{equation}
  \label{eq:Qhk}
  Q\hk[x(\tau)] \equiv \mu^{-1}\IInt{\tau}{t_0}{t} \dot x(\tau)\vs(x(\tau))
\end{equation}
along a single stochastic trajectory $x(\tau)$. By comparing this expression
with eq.~\eqref{eq:fdt:v}, a connection between energy dissipation and the
violation of the \fdt\ for the velocity becomes obvious. Indeed, in
ref.~\cite{spec05a} we calculated the mean dissipation rate as
\begin{equation}
  \sigma \equiv \partial_t\mean{Q\hk} = \mu^{-1}\mean{\vs^2},
\end{equation}
which is constant in a \ness. For large time differences we get
\begin{equation}
  \mean{\dot x(t)\vs(x(\tau))} \xrightarrow{|t-\tau|\rightarrow\infty} 
  \mean{\vs}^2 = \vm^2.
\end{equation}
Since $\mean{\vs^2}\geqslant\mean{\vs}^2$, it follows that the violation of
the FDT is bounded by the energy dissipation rate,
\begin{equation}
  \label{eq:bound}
  \mean{\dot x(t)\vs(x(\tau))} \leqslant \mu\sigma,
\end{equation}
where the equal sign holds for $t=\tau$. This inequality specializes the
former derivation of a general upper bound for the FDT violation~\cite{cugl97}
to the present case of a driven colloidal particle.

It is interesting to note that we can equate the left hand side of
eq.~\eqref{eq:bound} involving two different times with the mean dissipation
rate $\sigma$ by taking into account the change of the generalized potential
$\Delta\phi\equiv\phi(x(t))-\phi(x(\tau))$ during the time $t-\tau$. With
\begin{equation}
  \frac{\vs(x(t))}{\vs(x(\tau))} = \frac{\ps(x(\tau))}{\ps(x(t))} 
  = e^{\Delta\phi}
\end{equation}
we see that
\begin{equation}
  \mean{\dot x(t)\vs(x(\tau))e^{\Delta\phi}} = \mu\sigma
\end{equation}
holds. Since in a \ness\ the total heat flow into the heat reservoir is
$Q\tot=Q\hk-T\Delta\phi$~\cite{hata01,spec05a}, the term $-T\Delta\phi$ can be
regarded as ``excess heat'' exchanged with the reservoir in addition to the
housekeeping heat. Its mean value $\mean{\Delta\phi}=0$ vanishes and it is
bounded, whereas $Q\hk$ increases in time on average.

\begin{figure}
  \twofigures[width=7.25cm]{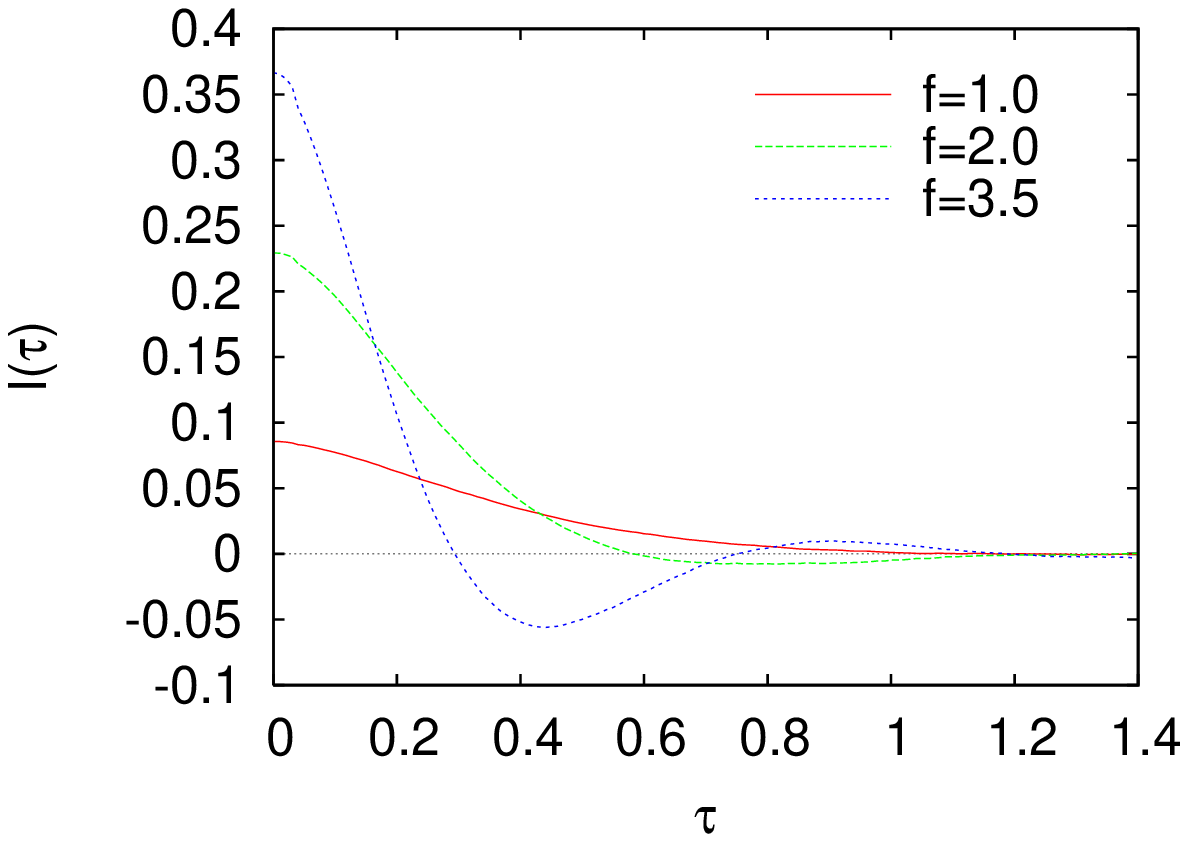}{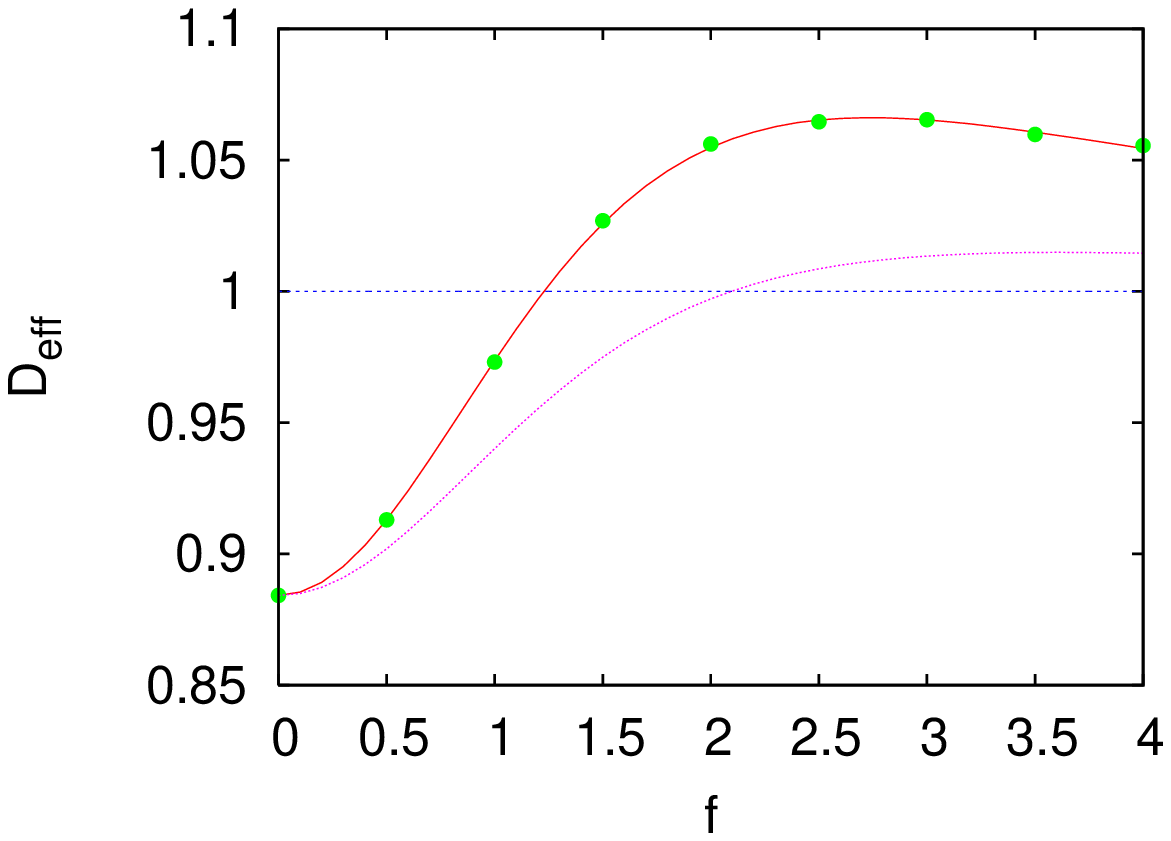}
  \caption{Normalized violation function $I(\tau)$ over time difference $\tau$
    obtained from 250,000 simulated Langevin trajectories for the periodic
    model potential $V(x)=V_0\cos2x$ (with $V_0=\frac{1}{2}$, $D=T=1$). For
    large forces $f$, the function $I(\tau)$ shows a non-monotonic decay.}
  \label{fig:corr}
  \caption{The effective diffusion constant $\Deff$ over the force $f$ for the
    same potential and parameters as in fig.~\ref{fig:corr}. The solid line
    shows the analytical solution~\eqref{eq:deff:exact} whereas the closed
    circles are obtained numerically from eq.~\eqref{eq:deff}. The dotted line
    shows the susceptibility~\eqref{eq:susc}.}
  \label{fig:diff}
\end{figure}

\section{Perspectives}

A direct experimental test of the violated or restored FDT,
eqs.~\eqref{eq:fdt:v} or~\eqref{eq:fdt:claim} respectively, will certainly be
a challenge. We therefore suggest first to test an integrated version
involving the effective diffusion coefficient. As a transport coefficient, it
is related to velocity correlations by the Green-Kubo formula
\begin{equation}
  \Deff \equiv \lim_{t\rightarrow\infty}\frac{\mean{x^2(t)}-\mean{x(t)}^2}{2t}
  = \IInt{\tau}{0}{\infty} \mean{[\dot x(\tau)-\vm][\dot x(0)-\vm]}.
\end{equation}
Inserting eq.~\eqref{eq:fdt:v} leads to
\begin{equation}
  \label{eq:deff}
  \Deff = T\left.\pd{\mean{\dot x}\p}{\fpr}\right|_{\fpr=0}
  + \IInt{\tau}{0}{\infty}I(\tau),
\end{equation}
where the first term on the right hand side is the static susceptibility
\begin{equation}
  \label{eq:susc}
  \IInt{\tau}{0}{\infty}R_{\dot x}(\tau)
  = \left.\pd{\mean{\dot x}\p}{\fpr}\right|_{\fpr=0}
\end{equation}
and the second term the integral over the normalized violation function
$I(t)$. All three quantities appearing in eq.~\eqref{eq:deff} can be measured
independently.

For a numerical check, we calculate the mean velocity in eq.~\eqref{eq:susc} as
\begin{equation}
  \mean{\dot x}\p = \frac{1-\exp[-l(f+\fpr)/T]}{\IInt{x}{0}{l} I_+(x)/l},
  \quad
  I_\pm(x) \equiv
  \frac{1}{D}\IInt{z}{0}{l}\exp\left\{\pm[U(x)-U(x\mp z)]/T\right\}
\end{equation}
with $U(x)\equiv V(x)-(f+\fpr)x$ and the effective diffusion coefficient as
\begin{equation}
  \label{eq:deff:exact}
  \Deff = Dl^2\left[\IInt{x}{0}{l} I_+(x)\right]^{-3}
  \IInt{x}{0}{l} I_+^2(x)I_-(x),
\end{equation}
where now $\fpr=0$~\cite{reim01}. The violation function $I(t)$ as shown in
fig.~\ref{fig:corr} is calculated from simulated Langevin trajectories. In
fig.~\ref{fig:diff}, we compare the known expression~\eqref{eq:deff:exact}
with the numerical results obtained from eq.~\eqref{eq:deff} with excellent
agreement.

Although we discuss in this Letter the one-dimensional case, following our
route it is clear how to generalize eq.~\eqref{eq:fdt:xdot} to interacting
systems with more than one degree of freedom. Furthermore,
eq.~\eqref{eq:fdt:claim} remains valid even if we drop the restriction of the
Einstein relation in favor of an ``effective'' temperature $\Teff\equiv
D(f)/\mu$, where the strength of the noise $D(f)$ depends on the driving.
Whether our approach to restore a FDT by referring the velocity to the
appropriate local mean velocity can be extended to non-velocity like
observables involved in violations of FDT's remains to be seen.

\acknowledgments

Valuable comments by A. Gambassi and an anonymous referee are gratefully
acknowledged.


\end{document}